%% file: emnlp2022.tex
\pdfoutput=1

\documentclass[11pt]{article}

\usepackage{EMNLP2022}

\usepackage{times}
\usepackage{latexsym}

\usepackage[T1]{fontenc}

\usepackage[utf8]{inputenc}

\usepackage{microtype}

\usepackage{inconsolata}

\usepackage{amsmath,amssymb,amsfonts}
\usepackage[vlined,ruled]{algorithm2e}
\usepackage{graphicx}
\usepackage{textcomp}
\usepackage{xcolor}
\usepackage{dutchcal}
\usepackage{enumitem}
\usepackage{balance}
\usepackage{multirow}
\DeclareMathOperator*{\argmax}{argmax}

\title{Can Visual Context Improve Automatic Speech Recognition \\for an Embodied Agent?}

\author{Pradip Pramanick \and Chayan Sarkar\\
    Robotics \& Autonomous Systems \\ TCS Research, India \\
  \texttt{\{pradip.pramanick,sarkar.chayan\}@tcs.com} \\
  }

\begin{document}
\maketitle

\input{abstract}

\input{intro}
\input{background}
\input{system}

\input{algo}

\input{eval}

\input{conclusions}
\input{limitations}
\input{ack}



\balance

\bibliographystyle{acl_natbib}
\bibliography{main}

\end{document}

%% file: abstract.tex
The usage of automatic speech recognition (ASR) systems are becoming omnipresent ranging from personal assistant to chatbots, home, and industrial automation systems, etc. Modern robots are also equipped with ASR capabilities for interacting with humans as speech is the most natural interaction modality. However, ASR in robots faces additional challenges as compared to a personal assistant. Being an embodied agent, a robot must recognize the physical entities around it and therefore reliably recognize the speech containing the description of such entities. However, current ASR systems are often unable to do so due to limitations in ASR training, such as generic datasets and open-vocabulary modeling. Also, adverse conditions during inference, such as noise, accented, and far-field speech makes the transcription inaccurate. In this work, we present a method to incorporate a robot's visual information into an ASR system and improve the recognition of a spoken utterance containing a visible entity. Specifically, we propose a new decoder biasing technique to incorporate the visual context while ensuring the ASR output does not degrade for incorrect context. We achieve a 59\% relative reduction in WER from an unmodified ASR system.

%% file: intro.tex
\section{Introduction}
\label{sec:intro}
Spoken interaction with a robot not only increases its usability and acceptability, it provides a natural mode of interaction even for a novice user. The recent development of deep learning-based end-to-end automatic speech recognition (ASR) systems has achieved a very high accuracy~\citep{li2021recent} as compared to traditional ASR systems. As a result, we see a huge surge of speech-based interfaces for many systems including robots. However, the accuracy of any state-of-the-art ASR gets significantly impacted based on the dialect of the speaker, distance of the speaker from the microphone, ambient noise, etc., particularly for novel and low-frequency vocabularies. These factors are often predominant in many robotic applications. This not only results in poor translation accuracy but also impacts the instruction understanding and task execution capability of the robot. 

\begin{figure}
    \centering
    \includegraphics[width=\linewidth]{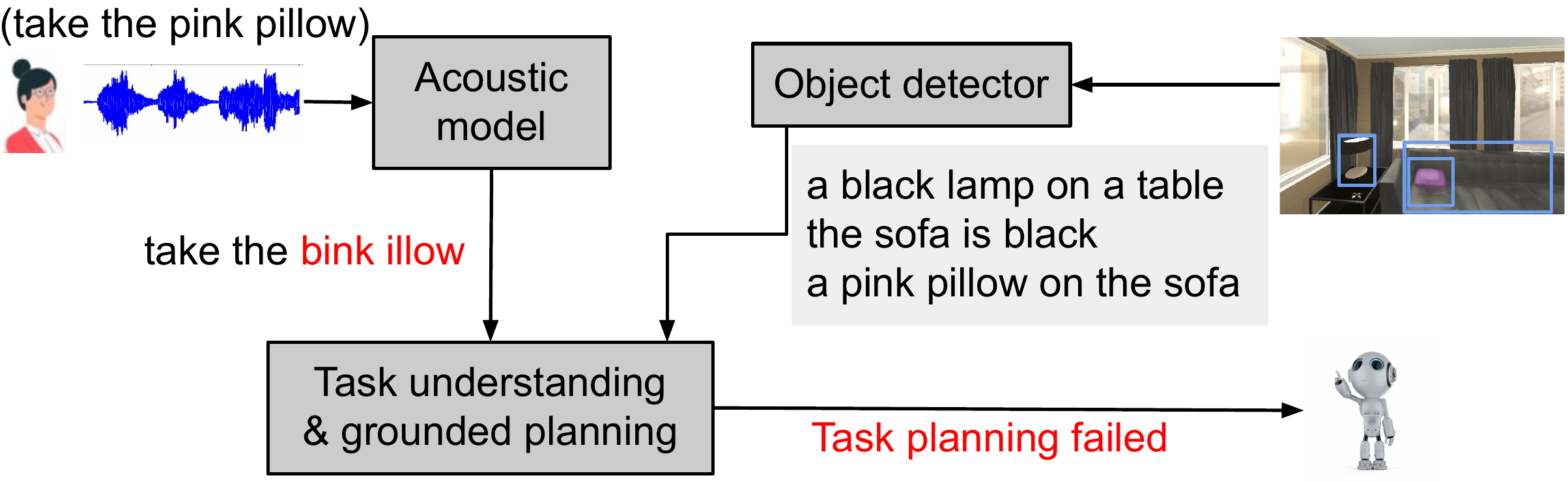}
    \caption{A simple pipeline of speech interface for human-robot interaction.}
    \label{fig:simple_pipeline}
    \includegraphics[width=\linewidth]{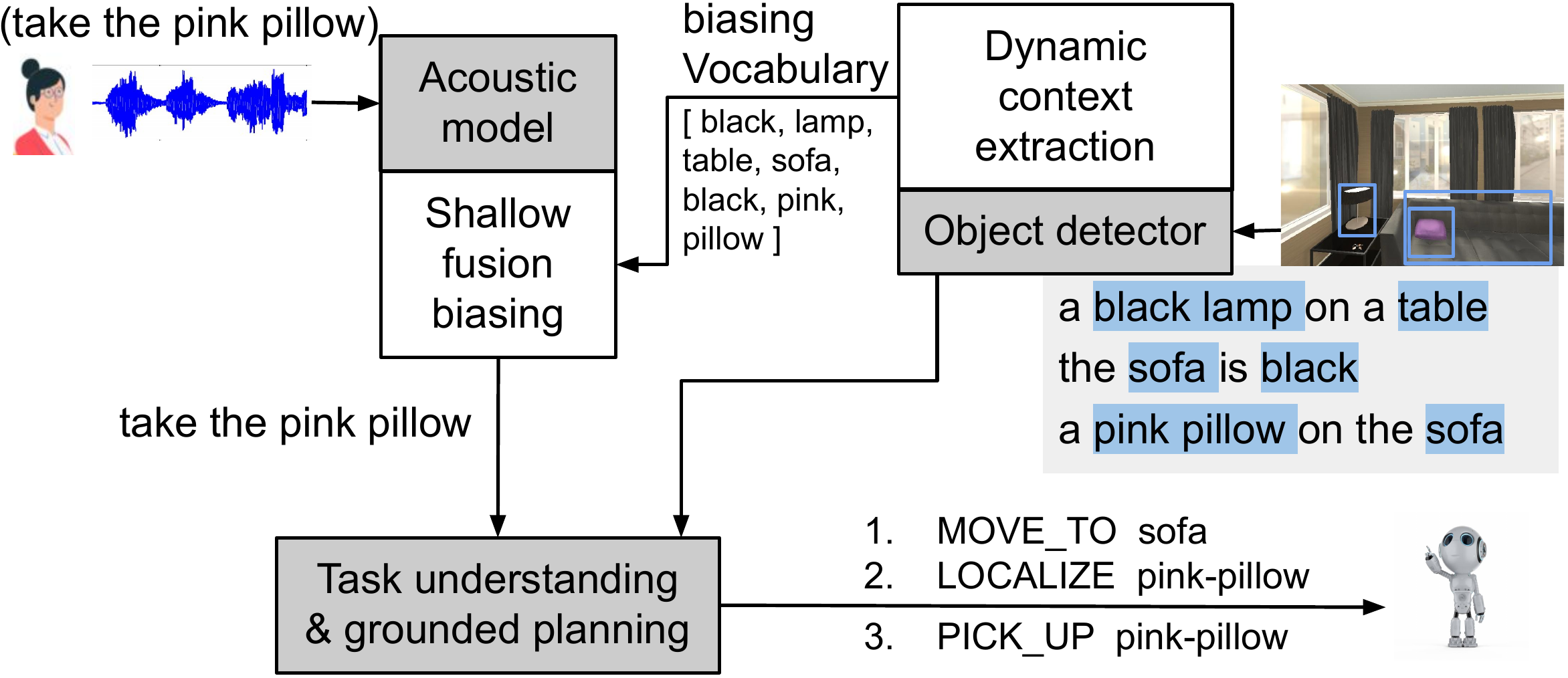}
    \caption{\textbf{Ro}bust \textbf{S}peech \textbf{I}nterface (\textbf{RoSI}) for embodied agents with shallow fusion biasing using dynamic biasing vocabulary.}
    \label{fig:rosi_pipeline}
    \vspace{-0.5cm}
\end{figure}

Figure~\ref{fig:simple_pipeline} depicts a typical scenario where an agent\footnote{In this article, we use robot and agent interchangeably.} uses an ASR to translate audio input to text. Then, it detects the set of objects in its vicinity using an object detector. Finally, it matches the object mentioned in the command and the objects detected in the vicinity (grounding) to narrow down the target object before execution. If the audio translation is erroneous, the grounding can fail, which leads to failure in task execution. For example, in Figure~\ref{fig:simple_pipeline}, even though the user mentioned ``pink pillow'', the translation was ``bink illow'', which results in failure in task grounding. 

There has been an increasing interest in contextual speech recognition, primarily applied to voice-based assistants~\citep{williams2018contextual,pundak2018deep,chen2019end,he2019streaming,gourav2021personalization,le2021contextualized}. However, incorporating visual context into a speech recognizer is usually modeled as a multi-modal speech recognition problem~\citep{michelsanti2021overview}, often simplified to lip-reading~\citep{ghorbani2021listen}. Attempts to utilize visual context in robotic agents also follow the same approach~\citep{ONEATA2021106943}. Such models always require a pair of speech and visual input, which fails to tackle cases where the visual context is irrelevant to the speech. 

In contrast, we consider the visual context as a source of dynamic prior knowledge. Thus, we bias the prediction of the speech recognizer to include information from the prior knowledge, provided some relevant information is found. There are two primary approaches to introducing bias in an ASR system, namely shallow and deep fusion. Shallow fusion based approaches perform rescoring of transcription hypotheses upon detection of biasing words during beam search~\citep{williams2018contextual,kannan2018analysis}. Class-based language models have been proposed to utilize the prefix context of biasing words~\citep{chen2019end,kang2020fast}. \citealt{zhao2019shallow} further improved the shallow-fusion biasing model by introducing sub-word biasing and prefix-based activation.~\citealt{gourav2021personalization} propose 2-pass language model rescoring with sub-word biasing for more effective shallow-fusion. 

Deep-fusion biasing approaches use a pre-set biasing vocabulary to encode biasing phrases into embeddings that are applied using an attention-based decoder~\citep{pundak2018deep}. This is further improved by using adversarial examples~\citep{alon2019contextual}, complex attention-modeling~\citep{chang2021context,sun2021tree}, and prefix disambiguation~\citep{9415054}. These approaches can handle irrelevant and empty contexts but are less scalable when applied to subword units~\citep{pundak2018deep}. Furthermore, a static biasing vocabulary is unsuitable for some applications, including the one described in this paper. Recent works propose hybrid systems, applying both shallow and deep fusion to achieve state-of-the-art results~\citep{le2021contextualized}. Spelling correction models are also included for additional accuracy gains~\citep{wang2021light,leng2021fastcorrect}.

In this article, we propose a robust speech interface pipeline for embodied agents, called RoSI, that augments existing ASR systems (Figure~\ref{fig:rosi_pipeline}). Using an object detector, a set of (natural language) phrases about the objects in the scene are generated. A biasing vocabulary is built using these generated captions on the go or it can be pre-computed whenever a robot moves to a new location. Our main contributions are twofold.
\begin{itemize}
    \item We propose a new shallow fusion biasing algorithm that also introduces a non-greedy pruning strategy to allow biasing at the word level using sub-word level information.
    \item We apply this biasing algorithm to develop a speech recognition system for a robot that uses the visual context of the robot to improve the accuracy of the speech recognizer.
\end{itemize}

%% file: background.tex
\section{Background}
We adopt a \textit{connectionist temporal classification} (CTC)~\citep{graves2006connectionist} based modeling in the baseline ASR model in our experiments. A CTC based ASR model outputs a sequence of probability distributions over the target vocabulary $\textbf{y}=\{y_1, \dots, y_T\}$ (usually characters), given an input speech signal with length $L$, $\textbf{x}=\{ x_1, \dots, x_L \}, L>T$, thus computing,
\begin{equation}
    P(\textbf{y}|\textbf{x}) = \prod_{i=1}^T P(y_i | \textbf{x}) .
\vspace{-0.35cm}
\end{equation}

The output sequence with the maximum likelihood is usually approximated using a beam search~\citep{hannun2014first}. During this beam search decoding, shallow-fusion biasing proposes rescoring an output sequence hypothesis containing one or more biasing words~\citep{hall2015composition,williams2018contextual,kannan2018analysis}. Assuming a list of biasing words/phrases is available before producing the transcription, a rescoring function provides a new score for the matching hypothesis that is either interpolated or used to boost the log probability of the output sequence hypothesis~\citep{williams2018contextual},
\begin{equation}
    s(y) = log P(y|x) - \lambda log B(y),
\vspace{-0.25cm}
\end{equation}
where $B(y)$ provides a contextual biasing score of the partial transcription $y$ and $\lambda$ is a scaling factor. 

A major limitation of this approach is ineffective biasing due to the early pruning of hypothesis. To enable open-vocabulary speech recognition, ASR networks generally  predict sub-word unit labels (e.g., character) instead of directly predicting the word sequence. However, as the beam search keeps a fixed number of candidates in each time-step $i \in L$, lower ranked hypotheses that contain incomplete biasing words, are pruned by the beam search before the word boundary is reached. 

To counter this, biasing at the sub-word units (grapheme/word-piece) by \textit{weight pushing} has been proposed in~\citep{pundak2018deep}. Biasing at the grapheme level improves recognition accuracy than word level biasing for speech containing bias terms, but the performance degrades severely for general speech where the bias context is irrelevant. Subsequently, two improvements are proposed - i) biasing at the word-piece level~\citep{chen2019end,zhao2019shallow,gourav2021personalization} and ii) utilizing known prefixes of the biasing terms to perform contextual biasing~\citep{zhao2019shallow,gourav2021personalization}. Although word-piece biasing shows less degradation than grapheme level on general speech, it performs worse than word level biasing when using high biasing weights~\citep{gourav2021personalization}. Also, prefix-context based biasing is ineffective when a biasing term is out of context. Moreover, a general problem with the weight-pushing based approach as compared to the sub-word level biasing is that they require a static/class-specific biasing vocabulary to work, usually compiled as a weighted finite state transducer (WFST). Also it requires costly operations to be included with the primary WFST-based decoder. However, for frequently changing biasing vocabulary, e.g., changing with the agent's movement, frequent re-compiling and merging of the WFST is inefficient.

Therefore, we propose an approach to retain the benefits of word-level biasing for general speech, also preventing early pruning of partially matching hypotheses using a modified beam search algorithm. During beam search, our algorithm allocates a fixed portion of the beam width to be influenced by sub-word level look-ahead, which does not affect the intrinsic ranking of the other hypotheses. This property is not guaranteed in weight-pushing~\citep{chen2019end,gourav2021personalization} that directly performs subword-level biasing. Moreover, we specifically target transcribing robotic instructions which usually include descriptions of everyday objects. Thus the words in the biasing vocabulary are often present in a standard language models (LM), while existing biasing models focus on biasing out-of-vocabulary (OOV) terms such as person names. We also utilize this distinction, by incorporating an n-gram language model to contextually scale the biasing score. We describe our shallow-fusion model in Section~\ref{sec:beam-search}.

%% file: system.tex
\section{System Overview}
\label{sec:system}
\begin{figure*}
    \centering
    \includegraphics[width=0.97\linewidth]{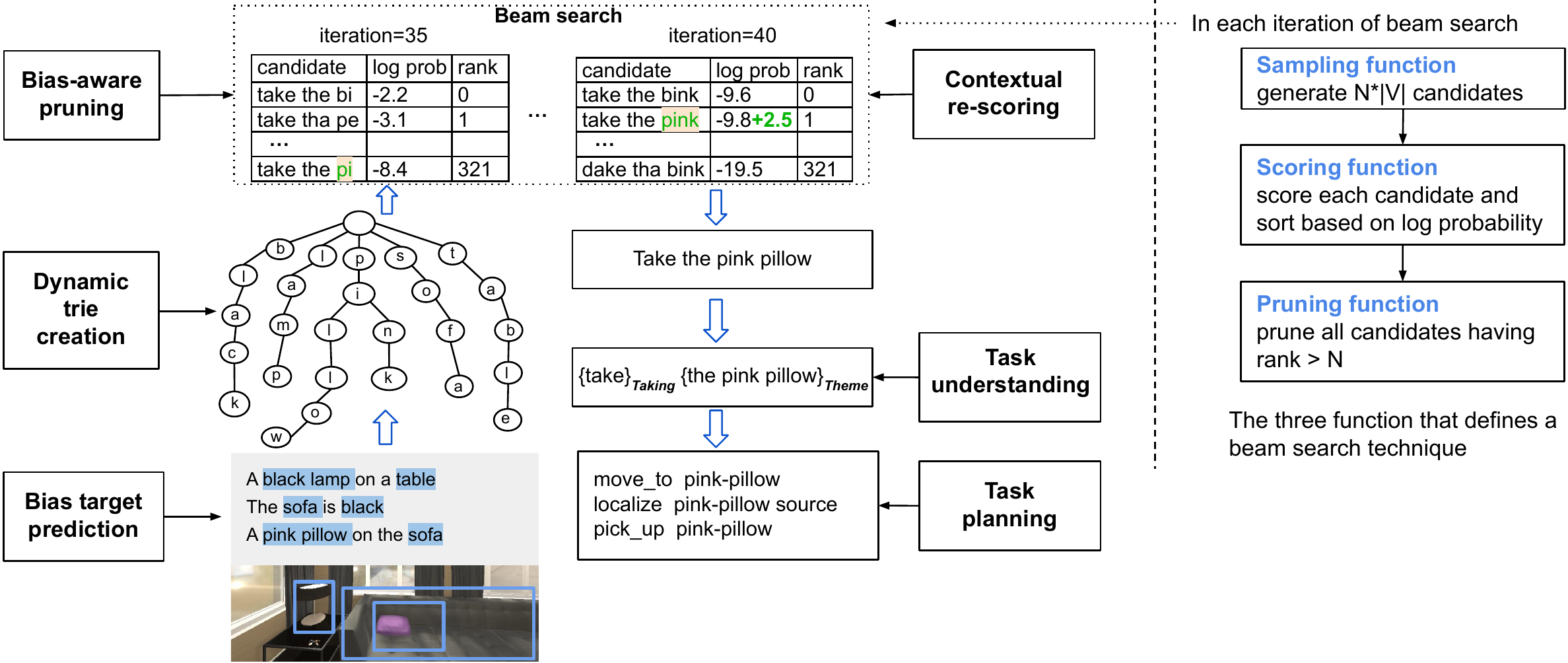}
    \caption{Our pipeline of robust speech interface (RoSI) for HRI that utilizes vision based dynamic contextual information. The functions that can be changed to change the beam search mechanism are highlighted.} 
    \label{fig:our_pipeline}
    \vspace{-0.5cm}
\end{figure*}

In this section, we present an overview of the embodied agent that executes natural language instructions, as depicted in Figure~\ref{fig:our_pipeline}. Given a speech input, the agent also captures an image from its ego-view camera. The \textit{dynamic context extraction} module extracts the visual context from the captured image before producing the transcription of the speech input. Firstly, a dense image captioning model predicts several bounding boxes of interest in the image and generates a natural language description for each of them. Given the dense captioned image, a bias target prediction model predicts a list of biasing words to be used in speech recognition. 

The list of biasing words/phrases is compiled into a prefix tree (trie) that is used by the beam search decoder to prevent the pruning of partially matched hypotheses. The trie is dynamically created with the agent's movement that captures a new image. The acoustic model processes the speech input to produce a sequence of probability distributions over a character vocabulary. We use the Wav2Vec2~\cite{NEURIPS2020_92d1e1eb} for acoustic modeling of the speech. This sequence is decoded into the transcription using a modified beam search decoding algorithm. During the beam search, the visual context that is represented using the biasing trie is used to produce a transcription that is likely to contain the word(s) from the visual context. We describe our biasing approach in detail in Section~\ref{sec:beam-search}. 

Given the transcribed instruction, the  \textit{task understanding \& planning} module performs task type classification, argument extraction, and task planning. We use conditional random field (CRF) based model as proposed in our earlier works~\cite{pramanick2019enabling, pramanick2020decomplex}. 
Specifically, the transcribed instruction is passed through a \textit{task-crf} model that labels tokens in the transcription from a set of task types. Given the output of task-crf, an \textit{argument-crf} model labels text spans in the instruction from a set of argument labels. This results in an annotated instruction such as,

\textit{[Take]\textsubscript{taking} [the pink pillow]\textsubscript{theme}}.

\noindent To perform a high-level task mentioned in the instruction (e.g., taking), the agent needs to perform a sequence of basic actions as produced by the task planner. The predicted task type is matched with a pre-condition and post-condition template, encoded in PDDL~\cite{mcdermott1998pddl}. The template is populated by the prediction of the argument-crf. Finally, a heuristic-search planner~\cite{hoffmann2001ff} generates the plan sequence. 


%% file: algo.tex
\section{Visual Context Biasing}
The probability distribution sequence produced by the acoustic model can be sub-optimally decoded in a greedy manner, i.e., performing an \textit{argmax} computation each time-step and concatenating the characters to produce the final transcription. However, a greedy-decoding strategy is likely to introduce errors in the transcription that can easily avoided by using beam search~\cite{NEURIPS2020_92d1e1eb}. In the following, we propose an approach to further reduce transcription errors by exploiting the embodied nature of the agent.

\subsection{Dynamic Context Extraction}
Upon receiving a speech input, the embodied agent captures an ego-view image. We process the image to identify the prior information about the environment that could be present in the speech. To do so, we detect all the objects in the image and generate a textual description for each. The object descriptions are further processed using a bias target prediction network to extract a dynamic biasing vocabulary. As the embodied agent performs a discrete action (such as moving or rotating), a new ego-view is captured, updating the visual context. Transcribing any new speech input after the action is executed, would be biased using the new context.

\subsubsection{Dense Captioning}
We utilize a dense image captioning network, DenseCap~\citep{yang2017dense} to generate rich referring expressions of the objects that includes self-attributes such as color, material, shape, etc., and various relational attributes. The DenseCap model uses a Faster R-CNN based region proposal network to generate arbitrarily shaped bounding boxes that are likely to contain objects-of-interest~\citep{yang2017dense}. The region features are produced by a convolutional network to predict object categories. The region features are further contextualized and fed into a recurrent network (LSTM) to generate descriptions of the proposed regions. We use a pre-trained model for our experiments. The model is trained on the Visual Genome dataset~\citep{krishna2017visual} containing approximately 100,000 real-world images with region annotations, making the model applicable to diverse scenarios.  

\subsubsection{Bias Target Prediction}
One could simply extract the tokens from the generated captions and consider the set of unique tokens (or n-grams) as the biasing vocabulary. However, a large biasing vocabulary could result in a performance degradation of the ASR system in case of irrelevant context, as shown in prior experiments~\citep{chen2019end,kang2020fast}. Therefore, we propose a more efficient context extraction approach, where we explicitly label the generated captions using the bias target prediction network.  Given a caption as sequence of word $\{ w_1, \dots, x_n \}$, the bias target prediction network predicts a sequence of labels $\{ l_1, \dots, l_n \}$ from the set of symbols \textit{S=\{B-B,I-B,O\}}, which denotes that the word is at the beginning, inside and outside of a biasing phrase. We model the  bias target predictor as a lightweight BiLSTM-based network. We encode the given word sequence using pre-trained GloVe embeddings~\citep{pennington2014glove}; thus producing an embedding sequence $\{ e_1, \dots, e_n \}$. We further obtain a hidden representation $h_i$ for each word $w_i$ by concatenating the two hidden representations produced by the forward and backword pass of the LSTM network. $h_i$ is fed to a feed-forward layer with softmax to produce a probability distribution over $S$,
\[  h_i= [ L\overrightarrow{S}TM(e_i,\overrightarrow{h}_{i-1}) ; L\overleftarrow{S}TM(e_i,\overleftarrow{h}_{i+1})   ]  \]
\[  l_{1:n} =\argmax_{l_i \in S} P(l_i | FNN(h_i)) .\]

\subsection{Beam Search Adaptation}
\label{sec:beam-search}
To enable word-level biasing while preventing early pruning of grapheme-level biasing candidates, we modify the generic beam search algorithm. More specifically, we modify the three general steps of beam search decoding as shown in Figure~\ref{fig:our_pipeline}. An overview of the modified beam search decoder is shown in Algorithm~\ref{algo:BAP_BSD-algo}. The algorithm accepts a sequence of pre-computed logits after applying softmax and returns the top-N (N is the beam width) transcriptions along with their log-probabilities. In the following, we describe its primary components in detail.
\subsubsection{Sampling function}
\label{sec:sampling}
In each time step of the beam search, existing hypotheses (partial transcription at time $t$) are extended with subword units from the vocabulary. Assuming the probability distribution over the vocabulary at time $t$ is $A_t$, this would generally result in the generation of $c_t=N \times |A_t|$ candidates, where $N$ is the beam width and $|A_t|$ is a constant denoting the dimension of the vocabulary. Please note that initially, i.e., at $t=1$, the hypothesis set is empty. Thus only $|A_1|$ candidates are generated, extending from beginning-of-sequence (<BOS>) tokens. The sampling function selects a subset of vocabulary to extend the hypotheses at time $t$ as,
\begin{equation}
    \mathcal{S}(A_t,C) = \{ a_t: \sum_{i=1}^{m} P(A_{t_i}) \approx C \},
    \label{sampling-eq}
\end{equation}
where $C$ is a hyper-parameter and $P(A_{t_i})$ is the probability of the $i^{th}$ index in the probability distribution at time $t$. Essentially, $\mathcal{S}$ starts pruning items from the vocabulary at time $t$ when the cumulative probability reaches $C$. Although a similar pruning strategy has been applied previously for decreasing decoding latency~\citet{amodei2016deep,NEURIPS2020_92d1e1eb}, we find that such a sampling strategy is essential to our biasing algorithm. By optimizing $C$ on development set, the decoder can be prevented from generating very low-scoring candidates, acting as a counter to over-biasing. This is particularly effective in irrelevant context, i.e., when none of the predicted biasing words are pronounced in the speech input.

\subsubsection{Contextual rescoring}
After the generation of candidates, the score of each candidate is computed in the log space according to the CTC decoding objective~\cite{hannun2014first}. We develop a rescoring function $\mathcal{R}$ that modifies the previously computed sequence score according to pre-set constraints. We apply $\mathcal{R}$ at the word boundary in each candidate, defined in the following,
\begin{equation}
\small
   \mathcal{R}(S_c) = 
    \begin{cases}
    S_c - \lambda log P_{lm} (c_n) & c_n \in V, c_n \in B_T  \\
    S_c - \delta & c_n \notin V, c_n \notin B_T \\
    S_c + \gamma & c_n \notin V, c_n \in B_T \\
    \end{cases}
    \label{re-scoring-eq}
\end{equation}
where $V$ is a word-vocabulary obtained from the set of unigrams of a n-gram language model, $B_T$ is the dynamic biasing trie, $c_n$ is the rightmost token in a candidate $c$, and $\gamma,\delta$ are hyper-parameters. The term $c_n \in B_T$ denotes the rightmost token (word) in candidate $c$ is a complete path in $B_T$. $S_c$ is the log probability of the character sequence in $c$, which is approximated using the CTC decoding equation and interpolated using a n-gram word level language model (lm) score~\cite{hannun2014first},
\begin{equation}
    S_c = log (P_{CTC}(c|x) P_{lm}(c_n|c_{n-1,\dots,c_1})^\alpha |c|^\beta),
    \label{lm-interpolation}
\end{equation}
where $P_{CTC}$ is the sequence probability computed from the output of wav2vec2, $P_{lm}$ is the same n-gram LM and $|c|$ denotes the word count in candidate $c$. The parameter $\alpha$ is scaling factor and $\beta$ is a discounting factor to normalize the interpolation with the sequence length.

In eq.~\ref{re-scoring-eq}, the rescoring function modifies the base sequence score according to the availability of certain contextual information. When the candidate ends in a word that is not OOV, it is likely that the language model has already provided a contextual boosting score (eq. \ref{lm-interpolation}). Therefore we compute the biasing score boost by simply scaling the lm-interpolated score using the unconditional unigram score of $c_n$. 

The rescoring function is similar to the unigram model in~\cite{kang2020fast}, but with two important distinctions. Firstly, we remove the class-based language modeling and propose simply using the unigram log-probability of a word-based language model. Thus we do not require any class-based LM training. Secondly, we propose reducing the score of an OOV, which is not a biasing word by the parameter $\delta$. This is based on the assumption that any OOV proposed by the ASR is less likely to be correct if it is not already in the biasing vocabulary. While the ASR is still a open-vocabulary system, i.e., it can produce OOV words, we impose a soft, conditional lexicon constraint on the decoder. This is also different from previously proposed hard lexicon constraint, applied unconditionally~\cite{hannun2014first}. For the third condition in eq.~\ref{re-scoring-eq}, we simply boost the score of a OOV in biasing vocabulary by a fixed amount, i.e., $\gamma$, as we do not have a contextual score for the same.

\subsubsection{Bias-aware pruning}
As the rescoring function for word-level biasing is applied at the word boundary, candidates can be pruned early without bias being applied, which could be otherwise completed in a valid word from the trie and rescored accordingly. We attempt to prevent this by introducing a novel pruning strategy. We define a rescoring likelihood function $\psi$ that scores candidates to be pruned according to the beam width threshold. As shown in Algorithm~\ref{algo:BAP_BSD-algo}, we divide the candidates into a \textit{forward} and a \textit{prunable} set. The forward set represents the set of candidates ranked according to their sequence scores, after applying $\mathcal{R}$. 

\begin{algorithm}[!ht]
\small
\LinesNumbered
\SetAlgoLined \DontPrintSemicolon
\SetKwInOut{Input}{Input}\SetKwInOut{Output}{Output}
\SetKwInput{Initialize}{Initialization}
\Input{A (Probability distribution over alphabet) $*$ T (Time-steps) sized tensor, Beam width N, Hyper-parameter K, Biasing trie $B_T$.}
\Initialize{hypotheses=$\emptyset$}
\SetKwProg{alg}{Algorithm}
\alg{\algo}{
\For{$t \in T$}{
    {Sample $a_t$ from $A_t$ using sampling function $\mathcal{S}$} \;
    {Generate $N \times a_t$ candidates by extending each hypothesis} \;
    {Compute sequence log-probability of $N \times a_t$ candidates} \;
    \For{$c$ in candidates}{
        \If{$c$ ends a word boundary}
            {Rescore $c$ with rescoring function $\mathcal{R}$} \;
    }
    {Obtain H by sorting $N \times a_t$ candidates by descending score} \;
    {Initialize $forward=\{ c \in H, rank(c) <=N \}$} \;
    {Initialize $prunable= \{ c \in H, rank(c) > N \}$ } \;
    \For{$c \in prunable$ }{
            {Compute rescoring likelihood using $\psi$} \;
    }
    {Sort $prunable$ by descending rescoring likelihood} \;
    {Compute $k=K\%$ of $N$} \;
    {From $forward$, prune candidates having $rank(c)>(N-k)$} \;
    {Append top-$k$ candidates of $prunable$ to $forward$} \;
    {hypotheses $\leftarrow forward$} \;
}
\Output{Return top-N transcriptions}
}
\caption{Bias-aware pruning for beam search decoding.}
\label{algo:BAP_BSD-algo}
\end{algorithm}

The original beam search decoding algorithm makes a locally optimal decision (greedy) at this point to simply use the forward set as hypotheses for the next time-step and discard the candidates in \textit{prunable}. Instead, we formulate $\psi$, which has access to hypothetical, non-local information of future time-steps from $B_T$, to take a non-greedy decision. The rescoring likelihood (in log space) score calculates the probability of a candidate being rescored at a subsequent stage of the beam search. We approximate its value by the following interpolation,
\begin{equation}
    \psi(S_c) = S_c + \sigma log \Big( \frac{tn}{1+nl} \Big),
    \label{pruning-eq}
\end{equation}
where $tn$ (traversed nodes) is the nodes traversed so far in $B_T$, $nl$ (nodes to leaf) is the minimum no of nodes to reach a leaf node in $B_T$, and $\sigma$ is a scaling factor, optimized as a hyper-parameter. 

Essentially, we calculate the rescoring likelihood as a weighted factor of - i) how soon a rescoring decision can be made; which is further approximated by the ratio of character nodes traversed and minimum nodes left to complete a full word-path in $B_T$ and ii) the candidate's score which approximates the joint probability of the candidate's character sequence, given the audio input. In a special case when $\sigma$ is set to zero, $\psi$ simply represents the ranking by $S_c$. Thus, we compute the rescoring likelihood for candidates in the prunable set and sort in a descending order. Finally, we swap the bottom-$k$ candidates in the forward set with top-$k$ candidates in the prunable set. The rescoring likelihood is used only to select and prevent pruning of a subset of the candidates, but it doesn't change the sequence scores in any way. Thus, any rescoring due to bias is still applied at the word-level.

%% file: eval.tex
\section{Experiments}
\subsection{Data}
\label{sec:data}
To perform speech recognition experiments, we have collected a total of 1050 recordings of spoken instructions given to a robot. To collect the recordings, we first collect a total of 233 image-instruction pairs. The images are extracted from a photo-realistic robotic simulator~\cite{talbot2020benchbot} and two volunteers wrote the instruction for the robot for each given image. We recorded the written instructions spoken by three different speakers (one female, two males). All the volunteers are fluent but non-native English speakers. We additionally recorded the instructions using two different text-to-speech (TTS) models\footnote{https://github.com/mozilla/TTS}, producing speech as a natural female speaker. The male speaker and the TTS models each produced 233 recordings of instructions and the female speaker recorded 118 instructions. We divide the dataset into validation and test splits. The validation set contains 315 instruction recordings ($\approx 30\%$) and the test set contains 735 recordings.

\subsection{Baselines}
We compare our approach with several baselines as described in the following. All the baselines use a standard CTC beam search decoder implementation~\footnote{https://github.com/PaddlePaddle/PaddleSpeech}, with modified scoring functions (in bias-based models only) as described below.

\begin{itemize}[leftmargin=*]
    \item Base - This is a pre-trained wav2vec2 model that we use as the baseline acoustic model. Specifically, we use the \textit{wav2vec2-large} variant~\cite{NEURIPS2020_92d1e1eb}, fine-tuned on LibriSpeech~\cite{panayotov2015librispeech}.
    \item Base + LM - We use a 3-gram, word vocabulary, language model trained on LibriSpeech text~\cite{panayotov2015librispeech}. The language model probabilities are interpolated using eq.~\ref{lm-interpolation}.
    \item Base + WB - We use a \textbf{W}ord-level \textbf{B}iasing approach that rescores LM-interpolated scores on word boundaries. This is similar to the word-level biasing described in~\cite{williams2018contextual}, but we use a fixed boost for bias.
    \item Base + WB\textsuperscript{ctx} - We use \textbf{W}ord-level \textbf{B}ias with \textbf{c}on\textbf{t}e\textbf{x}tual rescoring using unigram log-probability from LM. This is similar to~\cite{kang2020fast}, but instead of class-based estimation we use word-based LM probability, i.e., by setting $\delta=0$ in eq.~\ref{re-scoring-eq}.
\end{itemize}

\subsection{Optimization}
We optimize all hyper-parameters (except for the beam width $N$) for our system and the baselines with the same validation set. We use a Bayesian optimization toolkit~\footnote{https://ax.dev/docs/bayesopt.html} and perform separate optimization experiments for the baselines and our model, with a word error rate (WER) minimization objective. We use the same bounds for the common parameters, set the same random seed for all the models, and run each optimization experiment for 50 trials. The optimal hyper-parameter values and the corresponding search spaces for our model are shown in Table~\ref{tab:optimal-params}.
\begin{table}
\small
    \centering
    \begin{tabular}{|c|c|c|}
    \hline
         \textbf{Parameter} &\textbf{Value} &\textbf{Search space} \\ \hline
         $N$ & 100 & ---\\ \hline
         $C$ & 0.991 & [0.96, 0.9999]\\ \hline
         $\lambda$  & 1.424 & [0.005, 2.9] \\ \hline
         $\delta$   & 10.33 & [0.1, 14.0] \\ \hline
         $\gamma$   & 13.31 & [0.1, 14.0] \\ \hline
         $\alpha$   & 0.788 & [0.005, 2.9] \\ \hline
         $\beta$    & 0.119 & [0.005, 3.9] \\ \hline
         $\sigma$   & 10.91 & [0.001, 14.0] \\ \hline
         $K$        & 24 & [1, 35]\\ \hline
    \end{tabular}
    \caption{Optimal hyper-parameter values from search spaces. $N$ isn't optimized and a standard value is used.}
    \label{tab:optimal-params}
\end{table}

\subsection{Main results}
We primarily use WER metric\footnote{Computed using: https://github.com/jitsi/jiwer} for evaluation. We also show a relative metric, namely word error rate reduction (WERR)~\citep{leng2021fastcorrect}. Additionally, we use a highly pessimistic metric, namely transcription accuracy (TA), which computes an exact match accuracy of the entire transcription. We show the results of the experiments on the test set in Table~\ref{tab:main-results}.
\begin{table}
\small
    \centering
    \begin{tabular}{|l|c|c|c|}
    \hline
    \textbf{Model}       &\textbf{WER} \textdownarrow &\textbf{WERR} \textuparrow &\textbf{TA} \textuparrow \\ \hline
    Base        & 20.83  & ---      & 53.47 \\ \hline
    Base + LM     & 18.29  & 12.19    & 53.07 \\  \hline
    Base + WB     & 15.55  & 25.35    & 68.57       \\ \hline
    Base + WB\textsuperscript{ctx}      & 15.38      & 26.16      &68.71      \\ \hline
    Ours    & 8.48    & 59.28        &73.81   \\ \hline
    
    \end{tabular}
     \caption{Speech recognition results compared to baseline systems.}
    \label{tab:main-results}
\end{table}

Without any modification, the base ASR system produces an WER around 21\% and TA around 54\%. Even though predicting every word correctly is not equally important for task prediction and grounding, still these numbers in general do not represent the expected accuracy for a practical ASR setup in a robot. The LM interpolation reduces the WER by 12\%, but TA is slightly decreased by 0.75\%. Using word-level bias (Base+WB) results in a significant reduction in WER (25\%) and improvement in TA (28\% relative). Compared to Base+LM, WERR in Base+WB is 15\% and TA improvement is around 30\% relative. This again shows that even word-level biasing can effectively improve the ASR's accuracy, provided the biasing vocabulary can be predicted correctly. The Base+WB\textsuperscript{ctx} model performs slightly better than Base+WB, the WERR improving 26\% compared to Base and 1\% compared to Base+WB. Improvement in TA w.r.t. Base+WB is also minimal, i.e., 0.2\%. To analyze this, we calculated the percentage of the biasing words that are OOV's in the test set, which is found to be 9.5\%. Thus, due to lack of many OOV-biasing terms, the fixed boost part of the rescoring model in Base+WB\textsuperscript{ctx} was not triggered as much and model could not significantly discriminate between the scores of OOV and non-OOV biasing terms.

Our beam search decoder achieves a WER of 8.48, significantly outperforming the best baseline, i.e., Base+WB\textsuperscript{ctx} in both WER (45\% relative WERR) and TA (7.4\% relative improvement) metrics. Compared to the unmodified ASR model, the improvements are substantial, 59.3\% in WERR and 38\% in TA. We perform more ablation experiments on our algorithm in the following Section. 
\subsubsection{Other ablation experiments}
\label{sec:abalation}
\paragraph{Sampling and pruning function}
We experiment with no sampling, i.e., setting $C=1$ in the sampling function $\mathcal{S}$ (eq.~\ref{sampling-eq}) and setting $\sigma=0$ in the rescoring likelihood equation (eq.~\ref{pruning-eq}). The results are listed in Table~\ref{tab:abalation}. The first experiment shows that the sampling function is effective in our biasing algorithm. Although without it, we achieve 47.29\% WERR from the Base model, which is much higher than other baselines. The second experiment shows that the rescoring likelihood formulation improves the biasing effect, 53.53\% vs. 59.28 \% WERR, and improves the TA metric by 2.5\%.
\begin{table}
    \small
    \centering
    \begin{tabular}{|l|c|c|c|}
    \hline
    \textbf{Biasing model}  &\textbf{WER} &\textbf{WERR} &\textbf{TA} \\ \hline
    Ours, $C=1$    & 10.98     & 47.29     & 71.84 \\ \hline
    Ours, $\sigma=0$ & 9.68      & 53.53     & 71.98 \\ \hline
    \end{tabular}
    \caption{Effect of modifying sampling and pruning strategies.}
    \label{tab:abalation}
\end{table}
\paragraph{Anti-context results}
Biasing models make a strong assumption of the biasing vocabulary being relevant to the speech, i.e., a subset of the biasing vocabulary is pronounced in the speech. While this is often true when giving instructions to an embodied agent, it is important to evaluate biasing models against adversarial examples to measure degradation due to biasing. In the anti-context setting, we deliberately remove all words from the dynamic biasing vocabulary that are present in the textual instruction. We show the the results of anti-context experiments in Table~\ref{tab:anti-context-result}.
\begin{table}
    \centering
    \small
    \begin{tabular}{|l|c|c|c|c|}
        \hline
         \textbf{Biasing model}  & \textbf{WER} & \textbf{WERR} &\textbf{WERR\textsuperscript{*}}   &\textbf{TA} \\ \hline
         WB     &20.58  &1.21 & -32.3  &52.38 \\ \hline
         WB\textsuperscript{ctx}  &20.96      &-0.62      &-36.3    &51.16 \\ \hline
         Ours, $C=1$   & 14.36 & 31.1       &-30.8  &61.51  \\ \hline
         Ours, $\sigma=0$   & 12.72 & 38.9  &-31.4  &62.86 \\ \hline
         Ours   &11.09 &46.8    &-30.8 &65.85 \\ \hline
    \end{tabular}
    \caption{Speech recognition results in anti-context setting. WERR is relative to the Base model and WERR\textsuperscript{*} is relative to corresponding row for valid-context results (Table~\ref{tab:main-results} and Table~\ref{tab:abalation}) for the same model.}
    \label{tab:anti-context-result}
\end{table}
We find that our model has the least relative degradation compared to previous results in valid contexts. Even though all the biasing models are applied at word-level, prior experiments has shown that using sub-word level biasing often results in much worse degradation~\citep{zhao2019shallow,gourav2021personalization}. Also, the relative degradation for all variants of our model are lower than the baselines. More importantly, the absolute WER and TA values for our model are still much better than the unmodified ASR and other baselines, even in incorrect context.
\subsubsection{Qualitative results}
We show some examples from the dataset in Table~\ref{tab:qualitative}. 
In the first example, the word \textit{red} is a homophone of \textit{read}, and \textit{refrigerator} has long sequence of characters. Our model correctly transcribes both the words by utilizing the correctly predicted visual context. In the second example, \textit{frisbee} is a challenging word for ASR, which is also transcribed correctly using the biasing vocabulary. In the third example, \textit{row} is incorrectly predicted as \textit{roar} by both models. As the caption generator~\citep{yang2017dense} can't predict abstract concepts, e.g., \textit{row}, consistently, the visual context could not be utilized by our model. In the last example, our model correctly predicts \textit{stack} but incorrectly predicts \textit{books} as \textit{book}, as the visual context suggests that \textit{book} is likely to be spoken.
\begin{table*}
    \centering
    \small
    \begin{tabular}{|l|l|}
    \hline
         \textbf{Instruction} & \textbf{Visual context} \\ \hline
         \textit{Reference:} bring me the red book on the refrigerator & \multirow{3}{4cm}{..., book, red, refrigerator, ...} \\
         \textit{Base:} bring me the \underline{read} book on the \underline{refrijoritor} & \\
         \textit{Ours:} bring me the red book on the refrigerator & \\ \hline
         \textit{Reference:} take the frisbee on the floor & \multirow{3}{4cm}{..., frisbee, floor, ...} \\
         \textit{Base:} take the \underline{friezbee} on the floor & \\
         \textit{Ours:} take the frisbee on the floor & \\ \hline
        \textit{Reference:} move to the row of windows & \multirow{3}{4cm}{..., window, ...} \\
        \textit{Base:} move to the \underline{roar} of windows & \\
        \textit{Ours:} move to the \underline{roar} of windows & \\ \hline
        \textit{Reference:} pick up the stack of books & \multirow{3}{4cm}{..., book, stack, ...} \\
        \textit{Base:} pick up the \underline{steck} of books & \\
        \textit{Ours:} pick up the stack of \underline{book} & \\ \hline
    \end{tabular}
    \caption{Examples of transcriptions and their reference instructions. The first two examples show successful transcriptions, while the last two examples are failure cases. The relevant words shown as the visual context are extracted from the agent's ego-view. An underlined word indicates an error.}
    \label{tab:qualitative}
\end{table*}

%% file: conclusions.tex
\section{Conclusion}
In this article, we have presented a method to utilize contextual information from an embodied agent's visual observation in its speech interface. In particular, we have designed a novel beam search decoding algorithm for efficient biasing of a speech recognition model using prior visual context. Our experiments show that our biasing approach improves the performance of a speech recognition model when applied to transcribing spoken instructions given to a robot. We also find that our approach shows less degradation than other approaches when the extracted visual context is irrelevant to the speech. Even in adversarial context, the accuracy of our system is well beyond the accuracy of the unmodified speech recognition model.

%% file: limitations.tex
\section*{Limitations}
We have proposed a shallow-fusion biasing system that can be used in an embodied agent for improving accuracy in recognition of speech targeted to the agent, using visual context acquired by the embodied nature of the agent. The system's efficacy is naturally limited by the assumption that the speech is relevant to the visual context and the context can be extracted accurately. The first factor is a general limitation of any biasing system. However, we show that our system does not degrade too much if the context is irrelevant to the speech. 

As for the problem of context extraction, the system would also be less effective if the object detector/caption generator can not correctly predict the object name or describe the object as a caption. One way to counter this would be to additionally use a static prior knowledge, e.g., the entire vocabulary of an object detector or a pre-collected semantic map, instead of dynamic visual context extraction. The context also need not be visual only, e.g., one could bias using verbs (go, pick, bring, etc.) commonly used in instructions. 

We designed the system, specifically the beam search algorithm to be particularly effective when applied to embodied agents. While parts of the algorithm such as the pruning function are applicable to generic biasing problems, some parts such as the conditional lexicon constraint and our experiments in general are mostly limited to the robotics domain. For example, the average biasing vocabulary size in the test dataset is 10.9, which is typical for the described scenario. But we did not extensively test the algorithm's scalability to very large lists of biasing words. 

%% file: ack.tex
\section*{Acknowledgments}
We thank the anonymous reviewers for their valuable feedback. We thank Ruchira Singh, Ruddra dev Roychoudhury and Sayan Paul for their help with the dataset creation. We are thankful to Swarnava Dey for having helpful discussions.